# Service-Oriented High Level Architecture[1]

*Wenguang WANG+*
*Wenguang YU*
*Qun LI*
*Weiping WANG*
*Xichun LIU*
College of Information Systems and Management
National University of Defense Technology
Changsha, Hunan Province 410073 China
+86-731-4573558
wgwangnudt@gmail.com, wg.yu@126.com
liqun, wangwp@nudt.edu.cn, olive_simple@sina.com

**ABSTRACT:** *Recently, service-oriented paradigm is emerging as a new pattern following process-oriented and object-oriented ones in systems analysis and software development. Service-oriented High Level Architecture (SOHLA) refers to the high level architecture (HLA) enabled by Service-Oriented Architecture (SOA) and Web Services etc. techniques which supports distributed interoperating services. The detailed comparisons between HLA and SOA are made to illustrate the importance of their combination. Then several key enhancements and changes of HLA Evolved Web Service API are introduced in comparison with native APIs, such as Federation Development and Execution Process, communication mechanisms, data encoding, session handling, testing environment and performance analysis. Some approaches are summarized including Web-Enabling HLA at the communication layer, HLA interface specification layer, federate interface layer and application layer. Finally the problems of current research are discussed, and the future directions are pointed out.*

## 1. Introduction

High Level Architecture (HLA) was originally proposed by US Department of Defense (DoD) in 1995 as a common simulation framework to support the interoperability and reusability of various simulation applications inside DoD. Since the need of simulation interoperability extends outside of the defense community, HLA was accepted as an IEEE standard in September 2000[2-5]. Up to now, HLA is widely used as an open standard to connect live, virtual and constructive simulations for acquisition, training and testing in defense community.

Service-Oriented Architecture (SOA)[6,7] was proposed by Gartner Group in 1996[8,9] as a service-oriented framework to promote the reusability and interoperability of heterogeneous systems based on various operating systems, development platforms, programming languages and middlewares. SOA is an enterprise-oriented conceptual framework and Web Services[10] is a technique-oriented implementation framework, which is the prevailing technique to implement SOA. Service-oriented paradigm is immerging as a new pattern following process-oriented and object-oriented ones in systems analysis and software development. Some new terms are springing up such as service-oriented science[11], service-oriented computing[12], service-oriented design[13], service-oriented modeling[14], service-oriented simulation[15,16], service-oriented system engineering (SOSE)[17], service-oriented software engineering[18] and service-oriented Internet[19]. At present, major computer corporations including IBM, Microsoft and Sun, standards organizations such as Organization for the Advancement of Structured Information Standards (OASIS), the World Wide Web Consortium (W3C) and Simulation Interoperability Standards Organization (SISO) and many programs (e.g. DoD Net-Centric

---
[1] Most work of this paper is originally reported in [1] IN CHINESE by the authors since a paper is not necessary "original" to qualify for SIW Workshop.

Enterprise Services) are moving into or supporting the new paradigm.

As two different architectures supporting interoperability and reusability, HLA and SOA have been developed respectively within the defense simulation community and the commercial enterprise community. Recently combining HLA with SOA, especially using SOA to extend the capability of M&S framework in simulation community has attracted increasing attention[20]. Driven by the extension of application scope, the development of new technology and the need of net-centric simulation in Global Information Grid (GIG)[21], many deficiencies on interoperability, extensibility and reusability of HLA have been revealed during the past decade. Meanwhile, as an IEEE standard, HLA needs to be reviewed and revised every five years. Therefore SISO, which is an IEEE standard development organization, established HLA Evolved Production Development Group (PDG) to revise HLA[22-24]. HLA Evolved Web Service API (WS API)[25-29] was developed as an important enhancement to extend the capability of HLA by SOA and Web Services. Based on the proposed concept of Service-oriented HLA, this paper summarizes the current research of Web-Enabling HLA and pointes out the future directions.

## 2. Definition of Service-Oriented HLA

Service-oriented HLA refers to the architecture enabled by SOA and Web Services etc. techniques which supports distributed interoperating services. According to the layers of HLA shown in Figure 1, Web enabling HLA can be implemented at four layers:

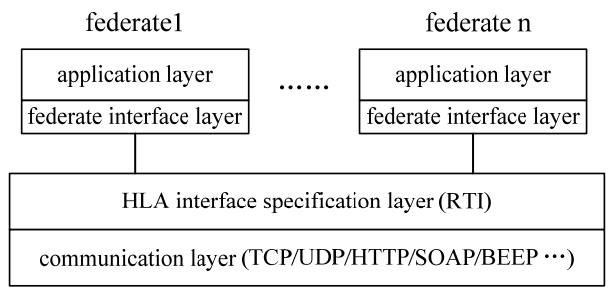

**Figure 1 Four layers of HLA**

- Web enabling HLA at the communication layer. The underlying communication protocol is enabled to support Web Services, for example, Web-Enabled RTI[30,31].

- Web enabling HLA at the interface specification layer. HLA interface specification is enabled to support Web Services. One approach is providing RTI as Web services such as HLA Evolved Web Service API and the other is interoperating directly with Web Service federates such as unified architecture[32].

- Web enabling HLA at federate interface layer. The interfaces between federates and RTI are enabled to integrate Web Service federates through adapters or connectors, for example, HLA Connector[32].

- Web enabling HLA at the application layer. The federates are enabled to become Web Service providers or clients such as HLA Island[28].

Both the interfaces in federate part and RTI part should conform to HLA interface specification, but the separation between them can separate the concerns of federates and RTI on one hand, and represent the partial order relationships of service providers/consumers and request/response mode on the other hand. To understand and apply service-oriented HLA, the comparisons between SOA and HLA and the importance of their combination are discussed in section 3. Then the current research on WS API is introduced in section 4, which is the latest progress in service-oriented HLA specification. In section 5 various approaches and applications of Web enabling HLA are discussed. The last section summarizes current research issues and points out the future directions of service-oriented HLA.

## 3. The comparisons between HLA and SOA

HLA is an interoperability-oriented integration architecture. The conceptual architecture of HLA is embodied in HLA framework and rules[2], Federation Development and Execution Process (FEDEP)[5] and the semantics of Object Model Template (OMT)[4], which is corresponding to SOA. The implementation architecture of HLA is embodied in HLA federate interface specification[3] and the syntax of OMT, which is corresponding to Web Services.

The comparisons between HLA and SOA, Web Services are shown in Table 1. The levels of interoperability are defined by Tolk's Levels of Conceptual Interoperability Model[33]. Reusability refers to the ability that components, models or systems can still be reused when the context of the application changes[34]. Referring to the definition from Defense Modeling and Simulation Office (DMSO) [35], we define extensibility as the ability that systems are able to add or remove new resources, application, equipments or subsystems over time and space without impacting their initial design. Scalability refers to the ability of a distributed simulation to maintain temporal and spatial consistency as the number of entities and accompanying interactions increase[35]. SOSE emphasizes on the specification, analysis, fault-tolerant computing, verification and validation, model checking, testing and dependability evaluation of service-oriented software and hardware[17].

**Table 1 Comparisons between HLA and SOA, Web Services**

| Comparison items | HLA | SOA and Web Services |
|---|---|---|
| Major application domain | Defense simulation community | Enterprise business community |
| Levels of interoperability | Syntax level | Semantic level can be reached by Web Ontology Language (OWL), semantic Web and business model |
| Interoperability | RTI implementation depends on computer platform. lower interoperability with non-M&S applications | Based on open and mature commercial standards and protocols, interoperability can be achieved by various applications following the standards |
| Reuse granularity | Simulation Object Model (SOM) and Base Object Model (BOM) | Coarse-grained services |
| Reusability | Implementation of federates and object models defined by SOM and BOM is dependent on computer hardware platforms, languages and middlewares | Services are described by WSDL. The open and standard interfaces eliminate heterogeneous underlying software and hardware platforms, languages and middlewares, leading to higher reusability |
| Reuse mode | One federate instance could only join one federation each time | Services could be reused by several applications simultaneously at any time |
| Extensibility | Constrained by FOM, extensibility is hard to be achieved in WAN and long distance connection, limited capability of integrating non-M&S tools and applications | Resources, applications and equipments in Internet or Intranet are easy to be integrated and supported by many corporations, development environments and software products |
| Scalability | Good scalability of interaction and information exchange in shared complex scenario | Lack of a common information exchange model, P2P connection mode may lead to lower scalability as the numbers of nodes and interactions increase |
| Coupling | Tight coupling among SOM, BOM and FOM | Loose coupling, indirect addressing. services are stateless and independent on contexts |
| Agility | Static architecture. Federates and interactions are fixed and rigid in FOM at integration time. The architecture is hard to change dynamically and recompose its components on demand at runtime. | Services have the properties of self-description and loose coupling. Higher agility can be achieved to implement a flexible framework by dynamic search, discovery and binding |
| Architecture style | Information bus style and having a shared information exchange model (FOM) | There are many ways to implement SOA. Web Services use P2P style and haven't a standard information exchange model |
| Systems Engineering Process | FEDEP emphasizes on interoperability but not the reusability of legacy federations and federates | SOSE emphasizes on reusability and collaboration of services |
| Performance | Higher performance in LAN | Lower performance in WAN, limited by network bandwidth |
| Time management, synchronization and states maintenance | Excellent. Providing the unique capability through logic clock to implement data exchange, synchronization and states maintenance between systems | Little support due to not specially designing for simulation |
| Publish, discover and indirect addressing | No, publish and subscribe according to FOM and SOM | Yes, having the specification of UDDI and directory service |
| Information exchange mode | Bidirectional call/callback interactions between federate and RTI and transferring data through TCP and UDP | Using unidirectional request/response interactions between client and server and transferring messages by SOAP and HTTP |
| Data encoding | Binary handle | XML and String |

We can draw the following conclusions from Table 1 that:

- HLA focuses on interoperability, interactions between components and effective information exchange among reusable resources. Reusability is the second goal and has not been solved very well.

- SOA and Web Services emphasize on reusability, loose coupling and the potential of resource components for reuse[34] but lack of a common information exchange model, and also have deficiencies at time management, synchronization, states maintenance and performance.

- Combining HLA with SOA, we can extend the interoperability, reusability, extensibility, agility etc. of HLA simulation framework by Web Services, at the same time keep inherit advantages of HLA in scalability, time management, synchronization, states maintenance and performance. Furthermore there are many needs of Web Services such as long distance connection, access through firewalls and integration of heterogeneous resources, by which the deployment and access of simulation application will be more convenient, so the combination of HLA and SOA and implementation by Web Services can largely extend the capability of M&S simulation framework.

## 4. Current research of Web Service API

The comparisons between HLA and SOA, Web Services illustrate the importance of their combination. HLA Evolved WS API is the latest progress to apply SOA idea to Web-enable HLA at interface specification layer. This section summarizes some issues about WS API from FEDEP, communication mechanisms, data encoding and so on.

### 4.1 The comparisons between WS API and native APIs

**Table 2 Differences between WS API and native APIs**

| Comparison items | WS API | Native APIs(such as C++, Java) |
| --- | --- | --- |
| Essence | Communication protocol between federates and RTI | Programming API |
| Relationship between RTI and federates | Production and consumption. WSPRC is located in server and federates in client. RTI may have several WSPRCs and each could provide services to several different federates at the same time in implementation. | Bidirectional call. RTI is responsible to management federates. RTI may have several LRCs and each could only connect one federate at the same time |
| Federate development and execution process | Web-centric thinking is throughout the whole process. Security, fault tolerance, update and interaction rate should be considered | No consideration about service-oriented or Web-centric thinking |
| Communication mechanism | Request/response, unidirectional | Call/callback, bidirectional |
| Transport protocol | HTTP | TCP or UDP |
| Encoding type | String and XML | Handle and binary |
| Session handling | Maintained by WSPRC using sessions | Maintained by LRC |
| Federate usage | Not need RTI lib. Service stubs are generated from WSDL | Need RTI lib |
| Development environment for federates and RTI | Languages and environments that support Web Services. WSDL could be mapped into various languages | Languages and environments compatible with the API |
| Application environment | WAN or LAN | LAN |
| Testing environment | Local or remote testing | Local or LAN testing |
| Performance | Lower and limited by bandwidth | Higher |

HLA WS API is also called WSDL API, by which simulation can be regarded as reusable services to interoperate with other Web Services in GIG environment. It has been a long time since the occurrence of Web-based simulation. In the late 1990s, many researchers realized the impact of Internet on M&S and proposed Web-based simulation concept[36-38], but at that time the use of Internet was limited only in long distance education and information acquisition. Later, Extensible Modeling and Simulation Framework

(XMSF)[39,40] was proposed to provide a common technology framework for both DoD and non-DoD M&S applications which uses the open and standard Web-based technologies to integrate M&S applications and operational systems on GIG. As a profile of XMSF, Web-Enabled RTI has been studied and applied in some projects[30,31]. HLA Evolved WS API enables HLA interface specifications, which can be seen as the further development of Web-Enabled RTI. Up to now, many commercial RTI corporations including Pitch and MAK are playing an active role in revising new HLA standard and developing new versions of RTI. Pitch has released his own commercial WS API product[41].

In this paper, native APIs refer to the ones in IEEE1516-2000 serial standards. Ada API is deleted from the original federate interface specification and WS API is added into the new HLA Evolved version. HLA WS API described by Web Services Description Language (WSDL, hla1516e.wsdl[42]) is the precise description of Web Services and not the one for programming. WSDL can be mapped into C++, Java, C# or VB so as to make use of various Web Services development frameworks and environments such as Apache Axis, Micrisoft.NET and IBM WebSphere to develop Web service federates. Corresponding to Local RTI Component (LRC) in native APIs, Web Service Provider RTI Component (WSPRC) is introduced into WS API, which changes the semantic of LRC from "fat client" to "thin client" and makes the developing, deploying and accessing RTI and Web service federates more convenient. The differences between HLA Evolved WS API and native C++/Java APIs are shown in Table 2 and the mapping from data structures of C++/Java API to Web Services is reported in [32]. But the enhancement of WS API is not to change the base semantic of HLA or redesign HLA. No matter which API or programming language a federate uses, it should be transparent to the federation and the other federates.

### 4.2 Federation Development and Execute Process

FEDEP provides systems engineering methodology guideline for HLA federation development and execute process. But the whole processes of service-oriented and Web-centric analysis, design and implementation are not considered in classical FEDEP. HLA Evolved WS API is favorable for federation developers to take advantages of Web Services from architecture to programming. Möller introduces WS API, reference FEDEP and some best practices for federation designers and programmers, including applying Web-centric thinking into federation design and execute process, establishing federation development and debug environment and considering some new issues in developing Web Service federates. The details are reported in [27].

### 4.3 Communication mechanisms

In native HLA APIs, ambassadors are responsible for communication between RTI and federate and their interactions are bidirectional with call/callback mechanism. On the contrary, in WS API, WSPRC and federate are respectively located in server and client and their interactions are unidirectional with request/response mode, which requires federates to call WSPRC EvokeMultipleCallbacks service at regular intervals to collect any pending callbacks and evoke additional calls[27]. One WSPRC may provide one or more ports, which is like an URL. One RTI may provide several WSPRCs and each could support several simultaneously connected federates. From transport protocol point of view, native APIs uses TCP unicast to transfer reliable data and uses UDP multicast to transfer other data with high update rate such as object attribute update. WS API is based on Web Services and transfers data with Simple Object Access Protocol (SOAP) and Hypertext Transfer Protocol (HTTP).

### 4.4 Data encoding

Native HLA APIs encode Federation Object Model (FOM) data with binary format and WS API with string format. So HLA Evolved WS API encodes object classes and attributes, interaction classes and parameters with string. To be compatible with native APIs, WS API provides several services to convert each other between handles and string names including "Get Object Class Handle"/"Get Object Class Name" [25]. Meanwhile, HLA Evolved introduces two standardized time representations of 64 bit float and 64 bit integer.

### 4.5 Session handling

Simulation models change their states over time but Web Services are inherent stateless, which brings difficulty to service-oriented simulation. Native APIs use LRC to maintain the states of local federates. Since there isn't any RTI ambassador in Web Service federates, WSPRC located in server makes use of sessions to maintain the states of remote federates.

The solution to session handling allows one application to join in one or more federations as several federates through several sessions. Additionally, it also allows several federates to connect to the same WSPRC or makes one federate play the role of "federation bridge" between federations. Therefore, HLA is enhanced by the reusability and flexibility of Web Services in this mechanism [32]. If the connection between WSPRC and federate breaks down due to federate or network failure, WSPRC will maintain the session for some time and if the connection is recovered in allowed period, simulation will resume which promotes fault-tolerance of simulation federation. Meanwhile, WSPRC checks sessions at regular times. The time-out sessions are treated as invalid ones and will be terminated to free WSPRC memory. Additionally, some new functions may be added in WSPRC, such as monitoring current sessions, diagnosing, logging or configuring parameters[27].

### 4.6 Testing environment

HLA Evolved WS API challenges traditional RTI testing environment and approaches. To establish a testing environment for it, McDonald et al. researched on the changes of HLA interfaces and modified RTI Verifier to support Web Services. Through the analysis of WS API communication mechanism, three kinds of verification strategies were proposed: local, manual remote and auto remote verification[29].

### 4.7 Performance analysis

Any system has to face trade-off between performance and modularization, reusability, interoperability. Although bringing advantages to M&S with loose coupling mechanism, WS API has lower performance than native APIs. In the "HLA WS API experimentation" carried out between US and Sweden, the performance of prototypical WS API in limited bandwidth WAN is equal to or lower than 3% of commercial Java API in gigabit LAN[26,32], and the movements of federates using Java API to exchange data are more smoothly than those using WS API. Möller holds that WS API will never achieve the same performance as commercial native APIs[32], but there are still a wide range of applications where this level of performance is adequate.

The main factors affecting the performance of WS API are listed as the following:

- Remote data transmission in WAN means more delay time, which is usually less than 1 millisecond in LAN may be hundreds of milliseconds or more in WAN.

- Limited bandwidth of WAN constrains transmission rate.

- Web Services is based on XML with string-encoded data and the encoding and decoding overhead makes transmission efficiency lower than that with binary data. Möller states that "XML calls have performance limitations that are highly caused by the XML encoding and decoding. Especially decoding a received XML message requires a lot of CPU. Performance in XML decoders needs to be considered to a higher degree. Another XML issue is that they cannot do multicast or best-effort so scalability is limited."[2]

- Web Services uses Client/Server architecture and request/response interaction mechanism, whose efficiency is lower than the bidirectional interaction mechanism in native APIs.

- If some additional mechanisms are considered in Web Service federates such as security, authentication and fault-tolerant, the overhead would lead to more delay.

Although WS API has lower performance than native APIs, it is still much valuable for the applications requiring lower temporal and spatial resolution, lower update rate and slower time advance rate, for example war game simulations such as Web-based Joint Theater Level Simulation (JTLS) [43].

The solutions to improving the performance of WS API are following: Firstly, in early design process of simulation systems, loose coupling services should be identified by the method of service-oriented system analysis. Subsystems with lower spatial and temporal resolution, lower update rate and slower time advance rate should be treated as Web Service federates. Secondly, the quantity and size of interaction messages should be reduced with XML compression technologies as much as possible. Thirdly, Dead Reckoning (DR) arithmetic should be used to make data prediction and extrapolation. Finally, message update rate should be regulated using HLA Evolved Smart Update Rate Reduction (SURR) [44] to avoid unnecessary data transmission. Meanwhile, the performance of WS API will be improved gradually with the upcoming commercial WS API products, the improvement of network performance and bandwidth and the occurrence of more effective message transportation mechanisms and protocols.

## 5. The approaches of Web Enabling HLA

HLA Evolved WS API Web-enables HLA at interface specification layer. Additionally, there are many other approaches of Web enabling HLA at the communication layer, federate interface layer and application layer. This section summarizes these approaches and applications at four different layers.

---

[2] Quote from Björn Möller's comments in discussion by email.

## 5.1 Web Enabling HLA at the communication layer

In this method, Web enables HLA underlying communication protocols and the typical example is Web-Enabled RTI[31], which uses Web-based communication protocols of SOAP and Blocks Extensible Exchange Protocol (BEEP) to communicate between HLA-compliant federates and RTI. In Web-Enabled RTI, federates reside on WAN as Web Services and end-users are able to use browsers to compose simulation federation. The advantages of this method are encapsulating non-reentrant RTI library, allowing several instances of Web Services provided by one federate on the same server and overcoming HTTP shortcomings of unidirectional call by BEEP, which supports bidirectional call and callback between federates and RTI. The disadvantages includes time representation inconsistency in different federates, the difficulty of encoding parameters with XML in Data Distribution Management (DDM) service, not supporting one Web Service federate to communicate with different RTIs simultaneously, the need of RTI and federate ambassadors stubs to be created manually and the requirement of developing Web Service wrapper to initiate federates remotely. The documents about Web-Enabled RTI Schema, definition of profiles and profiles prototype can be seen in [30,31,45].

This approach was used in Weapons of Mass Destruction Operational Analysis (WMDOA) simulation federation developed by Science Applications International Corporation (SAIC) for Defense Threat Reduction Agency (DTRA). In this project, DMSO RTI is able to communicate with SAIC RTI over Web by Web-Enabled RTI[31]. The WMDOA federation execute process model and Web-Enabled RTI architecture are shown in [31].

## 5.2 Web Enabling HLA at the interface specification layer

In this method Web enabling HLA is implemented at the HLA interface specification layer. One approach is that RTI is provided as Web Services and all the subsystems are treated as real HLA federates. The typical example is HLA Evolved WS API[28]. This approach combines SOA and HLA to provide the opportunity for simulation developers to take the advantages of Web Services and HLA. Developers could choose to use WS API, native APIs or their combination to develop various types of simulation systems according to different requirements [26].

This approach provides some obvious improvement: having a common information exchange model, supporting long distance communication, various operating systems and programming languages, flexible deployment and access, entirely supporting states maintenance, time management and synchronization and promoting interoperability between M&S and non-M&S applications. The deficiency is limited performance of Web Service part. This approach has been used in the project "HLA WS API experimentation" carried out by US and Sweden[32].

Another approach of Web enabling HLA interface specifications is Unified Architecture (UA)[32] proposed by Möller et al., shown in the Figure 2. The architecture has a United Infrastructure similar to RTI and is able to get a common information exchange model from the sum of services provided by HLA federates and SOA systems so as to facilitate interoperability under the Unified Architecture.

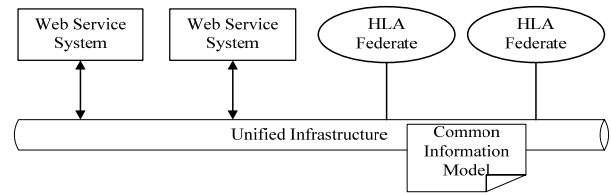

**Figure 2  Unified HLA/Web Service architecture[32]**

The Web services for the unified architecture can be regarded as the Cartesian product of FOM and HLA Interface Specification[32]:

$$UA\ Web\ Services = FOM \times HLA\ IFSPEC$$

This approach can map HLA into Web Services very well. But if each interaction, class and attribute in HLA is mapped into services, the granularity of Web Services will be very fine, which will lead to lower performance. In fact, the components and interactions in Web Services are coarse-grained, for example "check depot", while not like HLA to focus on updating the position attributes of an aircraft. The architecture is just a suggestion and no experimentation has been carried out.

## 5.3 Web Enabling HLA at the federate interface layer

This method enables the interfaces between federates and RTI to integrate Web Service federates through adapters or connectors such as HLA Connector [32]. The systems developed in SOA replace SOA Connector with HLA Connector and then they can interoperate and execute using HLA infrastructure [32]. Except the pros and cons HLA WS API has, the limitations of this method include that HLA Connector needs to be created manually and the update and interaction mapping from SOA services to HLA requires manual configuration.

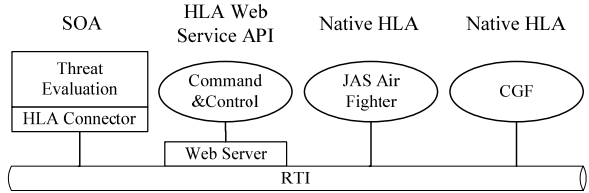

**Figure 3 Architecture of Swedish "HLA&SOA integration" in support for network-based defense[32]**

In the Swedish "HLA and SOA integration" in support for the network-based defense, the prototypical architecture has been implemented and tested that allows service-based and HLA-based systems to interoperate, shown in Figure 3. This project integrates four federates using the native API, WS API and HLA Connector respectively, which shows the feasibility of those approaches[32].

### 5.4 Web Enabling HLA at the application layer

In this method Web enabling HLA is carried out at the federate which on one hand participates in simulation, on the other hand acts as Web service providers or clients, for example, HLA Island [28]. This approach uses one federate to join HLA federation as a bridge to the other Web Service systems, providing Web Services or being a client to interoperate with other Web Service federates. The improvement of the approach is taking the advantages of HLA and Web Services. The limitations include that the federate as the bridge is application-dependent and needs to be created manually and the update and interaction mapping from Web Services to HLA cannot be established automatically, which leads to lower reusability.

### 5.5 Summary and Other approaches

Several approaches of Web enabling HLA are summarized in this section. At the communication layer, Web-Enabled RTI uses SOAP and BEEP to implement bidirectional communication between federates and RTI but is less flexible than WS API. At HLA interface specification layer, WS API could be supported by many Web Service development environments and software, but the performance of Web services is limited; the other approach Unified Architecture is an ideal infrastructure, but HLA needs to be redesigned with the thinking of SOA and needs further research and experiments. At federate interface layer, HLA Connector is the replacement of SOA Connector following the adapter principle but needs to be created manually. At the application layer, it is required to create the mapping between HLA and Web Services manually and the federate as the bridge is application-dependent, which leads to lower reusability.

In China, similar research has been done. Referring to XMSF framework, Huang Xiandong proposes a Web-based RTI platform architecture[46], which is similar to Web-Enabled RTI. Chen Xin et al. have implemented the management interface[47] between HLA and Web systems using COM, which corresponds to Web enabling HLA at the application layer. Han Chao et al. present several methods to extend HLA with Web Services[48]. In those methods, the extension at the federate layer is similar to Web enabling HLA at the application layer; the complete extension of RTI communication layer belongs to Web enabling at interface specification layer; the incompletely extension method is the mixture of Web enabling at the application layer and communication layer since the Web Service proxy federate is used as the bridge between the classic RTI and the Web-enabled RTI. Zhang Heming et al. propose a framework for collaborative M&S of complex systems, in which the mono-disciplinary simulation engines are encapsulated as Web Services and distributed collaborations between multi-disciplinary models are referred to HLA standards[49]. Jia Li et al. make domain models provided as Web Services and use HLA/RTI to integrate services[50]. The above two methods could be regarded as "HLA Island" and belong to Web enabling HLA at the application layer. Wang Hongwei et al. propose an approach to integrate multi-disciplinary models[51], which is similar to "HLA Island" and the federate as the HLA island plays the role of client to interact with Web Service federates. Xu Lijuan et al. present an HLA-based simulation service bus[52], which is similar to the idea of Unified Architecture at interface specification layer but requires coarse-grained services or lower data update rate and lower real-time services. The application mode and performance of this method need further research.

Finally, it should be further stated that this paper only summarizes the approaches of Web enabling the applications solely following HLA standards. The approaches of Web enabling HLA in other service-oriented frameworks such as OGSA are classified into the research issues of other frameworks, which is out of the scope of this paper.

## 6. Conclusion and Future Work

As the needs of simulation interoperability and reusability extend continually and Internet-based Web applications develop quickly, simulation architectures are developing towards standardization, hierarchization, networkitization, servicization, collaboratization and pervasivization[53,54]. The new service-oriented HLA combines the advantages of HLA and SOA and reflects the idea of simulation as services[55]. It provides a new solution to the integration, interoperability and reusability of heterogeneous resources and applications in simulation, enterprise and many other communities, which will promote the transformation of current simulation resources and the development of new applications, and have important research value and wide application prospect. From the needs of combing SOA with HLA, this paper particularly compares the two architectures, introduces WS API and summarizes the approaches of Web enabling HLA so as to provide the foundation for further research and application of service-oriented HLA.

Although the research on service-oriented HLA has made great progress, there are still many unsolved problems and difficulties which can be regarded as further research directions:

1) Research and development of service-oriented HLA itself

From the development of service-oriented HLA itself, unsolved issues include the interactions between coarse-grained services in SOA and fine-grained services in HLA, the research on the common information model between HLA and Web Services, the design and application of Unified Architecture, improving the performance of Web service federate, the implementation of Web Service standards (some development framework cannot create service stubs automatically according to WSDL), the interoperability in syntactic, semantic and conceptual levels and how to make full use of SOA (such as supporting simulation components for registering, finding and dynamic integration with UDDI).

2) Research on how to combine service-oriented HLA with other new technologies in HLA Evolved

From the development of HLA Evolved, some new improvements promote the composability, scalability, interoperability, reusability and reliability of HLA such as Base Object Model (BOM) [56,57], HLA Evolved modular FOM[58], XML Schema[23], HLA Evolved fault-tolerant federations[59], Dynamic Link Compatible APIs[23] and Smart Update Rate[44]. How to combine service-oriented HLA with these new technologies needs further research.

3) Research on interoperability and reusability between service-oriented HLA and other service-oriented M&S frameworks

From the development of service-oriented M&S framework, there are many different kinds of frameworks such as Discrete Event System Specification (DEVS) Unified Process Framework (DUNIP) [60], Dynamic Distributed Service-Oriented Modeling and Simulation Framework (DDSOS) [61], OGSA-Based (Open Grid Services Architecture) Simulation Framework[62], XMSF[39] and Service Integration/Interoperation Infrastructure (SI3) [63], which use various standards, specifications and technologies to implement service-oriented M&S framework. The relationship, interoperability and combination between these frameworks and service-oriented HLA are worth further researching.

## Acknowledgements

This work is supported by the National Natural Science Foundation of P.R. China (under grant Nos 60674069, 60574056). We thank Björn Möller, Per M. Gustavsson, Andreas Tolk, YANG Ruiping, CAO Xingping, HU Yanli and LEI Yonglin for the valuable discussions and constructive suggestions.

## Author Biographies

**WENGUANG WANG** is a Ph.D. student at National University of Defense Technology of China. His research interests are service-oriented simulation, simulation composability and interoperability, HLA, DEVS and system simulation. +Corresponding author Email: wgwangnudt@gmail.com

**WENGUANG YU** is a Ph.D. student at National University of Defense Technology of China. His research interests are system simulation and simulation-based evaluation.

**QUN LI** is an assistant professor at National University of Defense Technology of China. He got his Ph.D. in 1999 from National University of Defense Technology. His research interests are system simulation, simulation composability and simulation-based evaluation.

**WEIPING WANG** is a professor at National University of Defense Technology of China. He got his Ph.D. in 1997 from National University of Defense Technology. His research interests are system of systems engineering, simulation composability and simulation-based evaluation.

**XICHUN LIU** is a Ph.D. student at National University of Defense Technology of China. Her research interests are system simulation and simulation-based evaluation.